\documentstyle{elsart}

\def\simlt{\hbox{ \rlap{\raise 0.425ex\hbox{$<$}}\lower 0.65ex\hbox{$\sim$} }}
\def\simgt{\hbox{ \rlap{\raise 0.425ex\hbox{$>$}}\lower 0.65ex\hbox{$\sim$} }}

\def\that{{\hat t}}

\def\VEV#1{\left\langle #1\right\rangle}
\def\Nexp{N_{\rm exp}}

\def\pac{Paczy{\'n}ski}
\def\etal{{\it et al.}}

\def\that{{\widehat t}\,} 

\def\kpc{{\rm\, kpc}}

\def\msun {M_\odot}

\def\lsim{\mathrel{\mathpalette\@versim<}}
\def\gsim{\mathrel{\mathpalette\@versim>}}
\def\@versim#1#2{\lower0.2ex\vbox{\baselineskip\z@skip\lineskip\z@skip
  \lineskiplimit\z@\ialign{$\m@th#1\hfil##\hfil$\crcr#2\crcr\sim\crcr}}}
\def\spose#1{\hbox to 0pt{#1\hss}}
\def\simlt{\mathrel{\spose{\lower 3pt\hbox{$\mathchar"218$}}
     \raise 2.0pt\hbox{$\mathchar"13C$}}}
\def\simgt{\mathrel{\spose{\lower 3pt\hbox{$\mathchar"218$}}
     \raise 2.0pt\hbox{$\mathchar"13E$}}}

\begin{document}
\begin{frontmatter}
\title{Magellanic Cloud Gravitational Microlensing \\
       Results: What Do They Mean? }
\author{David Bennett}
\address{University of Notre Dame, Physics Department,
        \cty Notre Dame, IN 46556, \cny USA \\
        E-mail: bennett@nd.edu}

\begin{abstract}
Recent results from gravitational microlensing surveys of the Large Magellanic
Cloud are reviewed. The combined microlensing optical depth of the MACHO
and EROS-1 surveys is $\tau_{\rm LMC} = 2.1{+1.3\atop -0.8}\times 10^{-7}$
which is substantially larger than the background of 
$\tau_{\rm back} \leq 0.5\times 10^{-7}$ from lensing by known stellar
populations although it is below the expected microlensing optical depth of
$\tau = 4.7\times 10^{-7}$ for a halo composed entirely of Machos. 
The simplest interpretation of these results is that nearly half of the 
dark halo is composed of Machos with a typical mass of order $0.5\,\msun$.
This could be explained if these Machos are old white dwarfs, but
it is not obvious that the generation of stars that preceded these white
dwarfs could have gone undetected. It is also possible that Machos are
not made of baryons, but there is no compelling model for the formation of
non-baryonic Machos.
Therefore, a number of authors have been motivated to develop alternative
models which attempt to explain the LMC microlensing results with non-halo
populations. Many of these alternative models postulate previously unknown
dark stellar populations which contribute significantly to the total mass
of the Galaxy and are therefore simply variations of the dark matter solution.
However, models which postulate a unknown dwarf galaxy along the line of 
sight to the LMC or a distortion of the LMC which significantly enhances the
LMC self-lensing optical depth can potentially explain the LMC lensing
results with only a small amount of mass, so these can be regarded as
true non-dark matter solutions to the Macho puzzle. All such models
that have been proposed so far have serious problems, so there is, as yet,
no compelling alternative to the dark matter interpretation.
However, the problem can be solved observationally with a second generation
gravitational microlensing survey that is significantly more sensitive than
current microlensing surveys.
\end{abstract}

\end{frontmatter}
\section{Introduction}
In the past few years, the question of the composition of the Galactic
dark matter has changed from a topic of theoretical speculation to 
an experimental question. A number of different experiments have
or will soon have the sensitivity to probe most of the leading dark
matter candidates in realistic parameter regimes. 
The most exciting result to come from these
experiments so far has been the apparent detection of dark matter in the
form of Machos by the MACHO and EROS Collaborations \cite{LMC2,EROS1}. 
(Macho stands for MAssive Compact Halo Objects and refers to dark matter
candidates with masses in the planetary to stellar mass range.)
The MACHO
Project, in particular, has measured a gravitational microlensing optical
depth that greatly exceeds the predicted background of lensing by ordinary
stars. However, the timescales of the detected microlensing events indicate
an average lens mass that is firmly in the range of Hydrogen burning stars
or white dwarfs at $\sim 0.5\msun$. Both of these possibilities appear to
have other observable consequences that are not easy to reconcile with other
observations. This difficulty has led a number of authors to propose alternative
scenarios. One possibility is that the Machos are not baryonic despite
having similar masses to stars \cite{Jed,Yok}. Another possibility is that
there is some population of normal stars that contributes a much higher
microlensing optical depth than predicted by standard Galactic models.
In this scenario, the detected microlensing events are caused by
ordinary stars and have little to due with the Galactic dark matter.
This possibility is particularly
appealing to those searching for other dark matter candidates.

In this paper, we will discuss the microlensing technique as a dark matter
search tool, and we review the latest dark matter results from the MACHO
and EROS collaborations. We will then explore the various models that
have been proposed to explain these microlensing results, and we shall
see that there is no compelling alternative to the dark matter interpretation.
Then, we will show how observations of ``exotic" microlensing events
which do not conform to the standard microlensing lightcurve shape can
be used help pin down the location of the lens objects along the line of 
sight. Finally, we'll discuss the ``Next Generation Microlensing Survey"
which will have the sensitivity to discover a statistically significant
sample of ``exotic" microlensing events which should resolve the basic
question of whether the lensing objects reside in the Galactic halo.

\section{The Microlensing Technique}
Gravitational microlensing was proposed as a tool to study dark matter
in our galaxy by \pac\ in 1986
\cite{Pac}, and a few years later serious microlensing
dark matter searches were initiated \cite{RouRev,PacRev} 
These searches are sensitive to dark
matter in the mass range $10^{-7}-1\,\msun$, but region of greatest initial
interest was the $0.01-0.1\,\msun$ brown dwarf mass range.
\subsection{Microlensing Surveys of the Magellanic Clouds}
At present there are four gravitational microlensing surveys of the
Magellanic Clouds currently in operation.
The three original microlensing surveys, EROS, MACHO, and OGLE
all discovered their first microlensing events in September, 1993, and
all three are still in operation. MACHO was the first project to operate
a dedicated microlensing survey telescope with large format CCD cameras.
They have a 64 million pixel dual color imaging system on the 1.3m telescope
at Mt. Stromlo which has been observing for the microlensing project almost
exclusively since late 1992. To date, they've obtained over 70,000 images
of the Magellanic Clouds and the Galactic bulge. Each image covers about
half a square degree on the sky. Their most recently published dark 
matter results come from observations of 22 square degrees of the central
regions of the LMC over a period of 2 years ending in late 1994.

The EROS group is currently operating the EROS-2 telescope at La Silla.
This 1-m telescope is equipped with a 128 million pixel dual color camera 
which images 1 square degree-similar to the MACHO camera but with a field
of view that is a factor of 2 larger. Their published work is based upon
their earlier EROS-1 system which involved digitizing Schmidt Plates from
the ESO Schmidt telescope to search for microlensing events lasting
longer than a week as well as a CCD survey which imaged the 
central region of the LMC very frequently using a 0.4m telescope to
search for short timescale microlensing events.

The microlensing survey teams have yet to publish Magellanic Cloud
microlensing survey results. The OGLE group is now observing the clouds
with the OGLE-2 system which consists of a 1.3m telescope with a single
2k$\times$2k CCD camera operated in drift scan mode, but their published
survey results are from the OGLE-1 survey which did not observe the Clouds.
The other microlensing team observing the Magellanic Clouds is the 
MOA project which is a New Zealand/Japanese collaboration that observes the
LMC from a site in New Zealand that is far enough South that it is able to
observe the LMC at every hour of the night. 

\section{Microlensing Results from MACHO and EROS}
The first LMC microlensing events were discovered in September, 1993,
by the EROS and MACHO collaborations \cite{NatMACHO,NatEROS}. 
This was the first hint that
Machos might comprise a significant fraction of the Galactic Dark matter,
but the first quantitative measurement of a microlensing optical depth
significantly above the background didn't come until a few years later
when MACHO released the LMC results from their first two years of operation
\cite{LMC2}. They found eight microlensing events which indicated an
optical depth of 
$\tau_{\rm LMC} = 2.9{+1.4\atop -0.9}\times 10^{-7}$ which significantly
exceeds the expected microlensing optical depth of
$\tau_{\rm back} \leq 0.54\times 10^{-7}$ from known stellar
populations in the Galaxy and the LMC. Formally, the odds of obtaining
a measured optical depth as high as $\tau_{\rm LMC}$ in a Universe in
which the true microlensing is $\tau_{\rm back}  \leq 0.54\times 10^{-7}$
are about 0.04\%. A detailed comparison of the LMC
microlensing results from the MACHO Collaboration and the various
stellar lensing backgrounds is given in Table 1. The prediction for
a standard halo composed of Machos is also shown.

\begin{table}
\caption[]{Microlensing by Stars}
\label{tab-stars}
\begin{tabular*}{\hsize}{@{\extracolsep{\fill}}*{5}l@{}}
\hline
Population & $\tau (10^{-7})$ & $\VEV{\that}$ (days) &
$ \VEV{l} (\kpc)$  & $\Nexp$ \\
\hline
Thin disk   &  0.15  & 112  & 0.96  & 0.29  \\
Thick disk  & 0.036  & 105  & 3.0   & 0.075 \\
Spheroid    & 0.029  & 95   & 8.2   & 0.066 \\
LMC center  & (0.53) & 93   & 49.8  & (1.19) \\
LMC average & 0.32   & 93   & 49.8  & 0.71  \\ \hline
Total       & 0.54   & 99   &  -    & 1.14  \\ \hline
Observed    & $2.9{+1.4\atop -0.9}$ & 87   &  ?    & 8     \\ \hline
100\% MACHO dark halo   &  4.7   &  ?   & 14.4  &  ?    \\ \hline
\end{tabular*}\vspace{2pc}
\footnote{\ }{This table shows the predicted properties for microlensing 
 by known populations of stars for the MACHO 2-year data set.
 A Scalo PDMF is assumed, and the density and
 velocity distributions given in \cite{LMC1}. $\VEV{l}$ is the mean
 lens distance. The expected number of events $\Nexp$ includes the MACHO
 detection efficiency averaged over the $\that$ distribution.
 For the LMC, two rows are shown; firstly at the center, and
 secondly averaged over the location of our fields; only the
 averaged $\Nexp$ is relevant. For comparison, the observed values and
 those predicted for a halo composed entirely of MACHOs are shown.}
\end{table}

The EROS-1 microlensing survey has also reported an optical depth for the
LMC of $\tau_{\rm LMC} = 0.82 {+1.1\atop -0.5}\times 10^{-7}$ \cite{EROS1}. 
These error bars were not reported by the EROS collaboration, but they
were computing assuming Poisson statistics for events of the same
timescale. This value is substantially less than the MACHO 2-year result,
but because of the substantial statistical uncertainties, these have been
shown to be consistent with each other.
Because there is little overlap between the time and spatial
coverage of the MACHO 2-year and the EROS-1 experiments, the data sets are
essentially independent and it makes sense to average them. The relative
sensitivity of the two experiments can be determined by comparing the
expected number of events for the standard halo models for an assumed
Macho mass of $0.4\msun$. According to \cite{LMC2,EROS1}, 
EROS-1's sensitivity is about
58\% of that of the MACHO 2-year data set. With this weighting, we find
a combined EROS-1 \& MACHO-2-year microlensing optical depth of 
$\tau_{\rm LMC} = 2.1{+1.3\atop -0.8}\times 10^{-7}$.
\subsection{The Dark Matter Interpretation}
Since the MACHO and EROS experiments were designed as dark matter detection
experiments, the simplest interpretation of a signal above background is that
some of the dark matter halo has been detected. As shown in Table 1, the
line of sight to the LMC passes through only one known Galactic component
massive enough to provide a microlensing optical depth of 
$\tau_{\rm LMC} = 2.1{+1.3\atop -0.8}\times 10^{-7}$ and that is
the Galactic halo. However, the typical mass of the lenses is estimated to
be $\sim 0.5\msun$ which is well above the Hydrogen burning threshold. If
these $\sim 0.5\msun$ lens objects are made of ordinary Hydrogen and Helium,
then they would be bright main sequence stars which are far too bright to
be the dark matter.

A more reasonable choice for Macho dark matter would be white dwarf stars
which would be too faint to be easily seen. However, white dwarfs generally
form at the end of a star's life and only a fraction of the star's initial mass
ends up as a white dwarf. Thus, scenarios in which white dwarfs comprise
a significant fraction of the dark matter can be constrained by limits on
the brightness and heavy element production from the evolution of the
stars which preceded the white dwarfs \cite{Graff,Adams,Chab,Gib,Fields}.
However, a population of
white dwarfs which contributes substantially to the mass of the dark
halo must have some unusual characteristics in order to avoid appearing
as a normal old stellar population, so we must be wary of constraints
which rely upon the empirically determined properties of white dwarfs
as a hypothetical dark matter population would necessarily have some
differences.

Another possibility is that the Machos do make up the dark matter,
but they are not made of baryons. Non-baryonic Macho candidates include
such things as black holes, strange stars, and neutrino balls
\cite{Jed,Web}. These 
possibilities would behave as cold dark matter as far as galaxy formation
scenarios are concerned, so the non-baryonic Macho option would allow
Machos to comprise all of the dark matter in the Universe without
posing any difficulties for the galaxy formation scenarios favored by
cosmologists. It would seem to be a bit of a coincidence that the dark
matter mass would end up nearly the same as the mass of stars. One
scenario that might avoid this coincidence is the possibility of 
black hole formation at the QCD phase transition \cite{Jed} because
the horizon at the at the phase transition contains about a solar
mass of radiation. It takes only one horizon size region in 
$10^{8}$ to collapse to form a black hole in order to have
a critical density of black holes today. If the QCD transition
is first order, this might enhance horizon sized density 
perturbations somewhat, but this scenario probably requires
a feature in the density perturbation spectrum at the mass
scale of the observed Machos so it see that some sort of fine 
tuning is required in this scenario.

If the dark matter interpretation is correct, there is an additional
puzzle of the composition of the remainder of the dark matter since
the observed microlensing toward the LMC appears to be less than expected
for a dark halo made of only Machos. It could be that the rest of the halo
is made of particle dark matter such as WIMPs, axions, or massive neutrinos,
but another possibility is more massive Machos such as $\sim 20\msun$ black
holes which might form from the collapse of very massive stars. These events
would be 40 times less common than the observed shorter timescale events,
so current microlensing searches cannot put an interesting limit on them.
One microlensing event thought to be caused by a $\sim 20\msun$ black hole
has been seen towards the Galactic bulge, however.

\section{LMC Microlensing Events as a Background}
An alternative to the dark matter interpretation is the
possibility that the lensing background from ordinary
stars has been seriously underestimated. This requires
a substantial increase above the estimates shown in Table 1
which would require a substantial revision of the standard
Galactic or LMC model or else an entirely new component
of the Galaxy. The different models that have been proposed are
summarized in Table 2.
\subsection{Stars in the Galactic Disk or Spheroid}
The local Galactic disk is perhaps the best understood
part of the Milky Way, so it is difficult to make a modification
to the standard model of the disk of the magnitude required to
account for much of the observed microlensing signal. The thin
disk, in particular, is very well characterized. There are tight
constraints on the observed density of all types of stars in the
disk as well as constraints on the total column density of the
disk within 1 kpc of the Galactic plane: 
$\Sigma_0 < 80 \msun \,{\rm pc}^{-2}$. The observed density
of stars and gas in the thin disk comprises about half of this
limit \cite{GouldM}, so there is some room for additional material in
a massive thick disk. However, in order to explain the 
observed LMC microlensing events, the bulk of the matter in
this massive thick disk must be more than 1 kpc
from the Galactic plane. Furthermore, this thick disk is
far more massive than the thick disk that has been observed
in the stellar distribution, so it must be composed of objects
that are much darker than an ordinary stellar population. 

The Galactic spheroid is a roughly spherical distribution of stars with
a density that falls as $r^{-3.5}$ at large radii. This distribution has
been determined by comparison to star counts. If a standard
stellar mass function is assumed, then these data can be used to estimate
the microlensing optical depth yielding the result shown in Table 1.
In order to obtain a microlensing optical depth that can explain the
observed signal, a dark spheroid population with 50-100 times the mass
of the observed stellar population must be added.

Thus, both the spheroid and thick disk models are more properly 
considered to be variations of the dark matter interpretation rather
than lensing by the background of ordinary stars.
Because the thick disk and spheroid densities drop more rapidly at large 
radii than the canonical dark halo models, these
model predict a lower total mass in Machos for a given microlensing
optical depth than a standard halo model does \cite{Gates}.
This might ease some of the 
difficulties with the white dwarf lens option because these distributions
would require fewer white dwarfs in the Galaxy.

\subsection{Warped and Flared Disk: the Galactic Bagel}

A recent paper by Evans \etal \cite{bagel} has proposed that both
the thin and thick disk might be both flared and warped in
the direction of the LMC. This
seems like a reasonable option because warping is observed in
external galaxies and the gaseous component of our own
Galactic disk does appear to be flared. Interactions with
accreting dwarf galaxies like the recently discovered
Sagittarius dwarf might plausibly give rise to both
these effects. 

Evans \etal \cite{bagel} have investigated such
models and found that they did not yield an interesting
microlensing optical depth, so they turned to a more radical
set of models. The class of models that Evans \etal\ propose
has a disk column density that grows with Galactic radius
beyond the solar circle. Their most extreme model is
able to generate a microlensing optical depth of 
$\tau = 0.9\times 10^{-7}$ which is below the MACHO Project's 
95\% confidence level lower limit on $\tau_{LMC}$ even
when the other backgrounds listed in Table 1 are included.
However, some drastic assumptions are required in order to generate 
this relatively modest microlensing optical depth. This model has
a total disk mass of $1.5\times 10^{10} \msun$ inside of the
solar circle ($R_0 = 8\,$kpc), but it has $1.39\times 10^{11} \msun$
between 8 and $24\,$kpc. The total mass within $50\,$kpc is
$2.12\times 10^{11} \msun$.  Thus, the mass distribution resembles
that of a bagel with most of the mass contained within a toroidal
region outside of the solar circle. 

It is instructive to compare
this mass distribution to that of an isothermal sphere with
a circular rotation speed of $200\,$km/sec. The isothermal sphere has
a mass of $7.4\times 10^{10} \msun$ inside $8\,$kpc and
$1.49\times 10^{11} \msun$ between 8 and $24\,$kpc. (Note that a flattened
distribution of matter supports a faster rotation speed than a spherical
distribution.) Thus, Evans \etal's
Galactic bagel provides only a small fraction of the mass needed to
support the rotation at the solar circle, but it provides virtually
all of the mass needed between 8 and $24\,$kpc. Thus, if dark halo
provides the additional mass needed to support the Galactic rotation
curve, it must have a relatively high density inside the solar circle,
but essentially no mass between 8 and $24\,$kpc. The halo density would
grow once again to comprise most of the Galactic mass beyond $24\,$kpc.
Such a halo is virtually impossible with cold dark matter, but it might
be possible to make a halo that resembles this with massive neutrinos if
most of the mass inside the solar circle in a baryonic component that is
different from the Galactic disk/bagel.
In short, the Evans \etal\ model is similar to the thick disk model
discussed above in that the microlensing optical depth toward the
LMC is raised by significantly increasing the total stellar mass of the Galaxy.
It differs from the massive thick disk model in that the additional stars
are added far from the solar circle so that they can be brighter without
violating limits on the stellar content of the solar neighborhood.

\subsection{Magellanic Cloud Stars}

According to Table 1, lensing by ordinary stars in the LMC is the largest
contribution to microlensing background, and because the SMC is thought
to be elongated along the line of sight, the SMC self-lensing optical depth is
thought to be substantially larger than this \cite{EROSsmc}.
Furthermore, we don't have
as stringent constraints on star counts and the distance distribution of stars
in the Magellanic Clouds. These facts led Sahu and Wu \cite{Sahu,Wu} 
to suggest that perhaps lensing by LMC stars could be responsible for 
most of the observed LMC microlensing events. However, 
Gould \cite{Gouldself} showed that the self-lensing
of self gravitating disk galaxy, inclined by less than $45^\circ$ like the LMC,
is related to its line of sight velocity dispersion by the 
formula, $\tau < 4\VEV{v^2}/c^2$ under the assumptions that
the stellar disk is a relaxed virialized system. Since the velocity 
dispersion of the LMC is measured to be $20\,$km/sec, 
the implied self-lensing optical depth of the LMC is 
$\tau \simlt 2\times 10^{-8}$ which is far too low to explain the 
LMC microlensing events. 

This constraint can be avoided only if the LMC has a higher velocity 
dispersion than current measurements indicate - perhaps in the central
LMC where measurements are rather sparse and where the 
the microlensing search experiments have concentrated their
observations. Such a model would then predict that the microlensing
optical depth would be much lower in the outer LMC than in the
central bar. This is contrary to the observed distribution of event locations
which seems independent of distance from the LMC center, but the
current data sets are too small for a highly significant test of this
effect. Another possible way to evade Gould's limit on the LMC self-lensing
optical depth would be if the LMC is not a virialized self-gravitating
system.

\begin{table}
\caption[]{Models to Explain the LMC Microlensing Results}
\label{tab-models}
\begin{tabular*}{\hsize}{@{\extracolsep{\fill}}*{5}l@{}}
\hline
Lens Population & $\tau_{\rm pred}/\tau_{_{LMC}}$ & mass of pop. &
lens identity & problems \\
\hline
halo Machos  & 1  & $2\times 10^{11}\msun$ & WD or BH? & Macho formation? \\
dark thick disk & 1  & $\sim 10^{11}\msun$  & WD or BH? & Macho formation? \\
\ \ \& spheroid &    &   &  &  \\
foreground gal. & 1 & $10^9-10^{10}\msun$ & stars  & stars not seen; \\
                &   &                     &        & contrived \\
ZL foreground gal. & $<0.13$ & $10^9\msun$? & stars  & stars in LMC \\
warped flared disk & $\simlt 0.5$ & $\sim 10^{11}\msun$ & stars & contrived \\
LMC & $<0.2$ & $10^{10}\msun$ & stars & $\tau$ too small \\ \hline
\end{tabular*}\vspace{2pc}
\footnote{\ }{This table compares the various models that have been proposed
 to explain the LMC microlensing results.}
\end{table}

\subsection{Foreground Dwarf or Tidal Tail}

Inspired by the recent discovery of the Sagittarius Dwarf Galaxy
on the far side of the Galactic bulge, Zhao proposed that there
might be another similar sized dwarf galaxy along the line of
sight toward the LMC. This possibility could neatly explain the
LMC microlensing events with a normal stellar system of very
small mass, and so it involves no new population of Machos or
faint stars in unexpected locations. Like the Sagittarius Dwarf, 
such a galaxy could possibly have evaded detection because its
stars would generally be confused for LMC stars. The one obvious
drawback of this model is that such dwarf galaxies are quite
rare. The Sagittarius Dwarf is the only known dwarf galaxy with
a mass large enough to have a significant gravitational lensing
optical depth to explain the LMC microlensing results, and it
only covers about one thousandth of the sky. So, the {\it a priori}
probability of a chance foreground dwarf galaxy is only about 0.001.
Zhao \cite{Zhao} also suggested that a more likely scenario might be 
to have a ``tidal tail" of a Galaxy like the LMC or even the LMC
itself be responsible for the lensing events.

These ``foreground galaxy" models a significant boost when 
Zaritsky \& Lin \cite{ZL} (hereafter ZL)
reported the detection of a feature in the
LMC color-magnitude diagrams which they interpreted as
evidence for a population of ``red clump" stars in the foreground
of the LMC. However, neither Zhao's models or ZL's
interpretation of their observations has stood up very well under
scrutiny. The following counter arguments indicate that these
foreground galaxy or tidal tail models are not likely to be correct:
\begin{itemize}
  \item The MACHO Project \cite{RRLfore}
         showed that there is no excess population of
        foreground RR Lyrae stars toward the LMC indicating that there
        is no foreground dwarf galaxy with a old, metal poor stars
        at a distance of less than 35kpc towards the LMC.
  \item Beaulieu \& Sackett \cite{BS} argued that the CM diagram feature seen 
        by ZL is a feature of the LMC's giant branch rather than a foreground
        population.
  \item Bennett \cite{RCtau}
        showed that even if ZL's interpretation of the LMC CM
        diagram is correct, the implied microlensing optical depth is
        only 3-13\% of the microlensing optical depth seen by MACHO.
  \item Gould \cite{Gouldfore} showed that the outer surface brightness
        contours of the LMC do not allow for a foreground galaxy or tidal tail
        that extends beyond the edge of the LMC.
  \item Johnston \cite{Johnston}
        showed that tidal debris from the LMC or a similar
        galaxy would not be expected to remain in front of the LMC for
        any significant length of time.
\end{itemize}

\section{Exotic Microlensing Events}

Observations of exotic microlensing events such as caustic crossing 
events, parallax events, and binary source events can provide addition
information that can pin down the location of the lens. For example,
the recent observations of the binary caustic crossing for the
event MACHO Alert 98-SMC-1 have measured the time it takes for the
projected position of the lens center of mass to cross the diameter of
the source star \cite{EROSsmcb,PLANsmcb,MACHOsmcb}.
Since we can make a reasonable estimate of the source
star size from multicolor photometry or spectra of the source star,
a reasonable estimate of the angular velocity of the lens can be obtained.
In the case, of MACHO 98-SMC-1, the angular velocity was low indicating
that the lens resides in the SMC which is not a surprise because the
self-lensing optical depth of the SMC is expected to be large\cite{EROSsmc}.
Another
type of exotic event that has only been definitively observed towards
the galactic bulge is the microlensing parallax effect \cite{para} which is
a lightcurve deviation due to the motion of the Earth. This also yields
information on the distance to the lens although accurate photometry
is required to detect parallax effect for lenses in the halo. Also, the
converse of the parallax effect (sometimes called the xallarap effect)
can be observed if the source star is a binary and the effects of its
orbital motion can be seen. This requires multiple follow-up spectra to
characterize the binary source orbit \cite{HG}.

It is generally the case that the characterization of these exotic microlensing
events requires more frequent or higher accuracy photometry than can be
obtained by the microlensing survey teams, but this is now routinely
made possible by the routine discovery of microlensing events in real time
\cite{EWS,IAUC,Beck}. 
However, the rate that these exotic microlensing events are detected
in the LMC is only about 0.3 per year which is not enough to get a
statistically significant sample in a reasonable amount of time. However,
the next generation of microlensing surveys \cite{Stubbs}
should have a event detection
rate that is more than an order of magnitude higher than this which should
yield enough exotic events to resolve the Macho mystery.

\section*{Acknowledgment}   
This research was supported 
by the Center for Particle Astrophysics
through the Office of Science and Technology
Centers of NSF under cooperative agreement AST-8809616 and by the Lawrence
Livermore National Lab under DOE contract W7405-ENG-48.

\end{document}